\documentclass[authoryear]{elsarticle}

\usepackage[T1]{fontenc}
\usepackage[utf8]{inputenc}
\usepackage[british]{babel}
\usepackage{amsmath}
\usepackage{textcomp}
\usepackage{amssymb}
\usepackage{amsbsy}
\usepackage{url}
\usepackage{natbib}
\usepackage{graphicx}
\usepackage{caption}
\usepackage{subcaption}
\usepackage{hyperref}
\usepackage{cleveref}
\usepackage{stmaryrd}
\usepackage{sparklines}
\usepackage{booktabs}
\usepackage{xspace}
\usepackage{bm}

\graphicspath{{./Figs/}}

\newcommand{\x}{$\times$}
\newcommand{\eg}{e.g.\xspace}
\newcommand{\ie}{i.e.\xspace}

\renewcommand{\v}[1]{\ensuremath{\bm{#1}}} 
\newcommand{\gv}[1]{\ensuremath{\mbox{\boldmath$ #1 $}}}  
\newcommand{\grad}[1]{\gv{\nabla} #1} 
\renewcommand{\div}[1]{\gv{\nabla} \cdot #1} 
\newcommand{\pd}[2]{\frac{\partial #1}{\partial #2}} 

\DeclareMathAlphabet{\mathsfit}{\encodingdefault}{\sfdefault}{m}{sl}
\SetMathAlphabet{\mathsfit}{bold}{\encodingdefault}{\sfdefault}{bx}{sl}
\newcommand{\tensor}[1]{\bm{\mathsfit{#1}}}

\crefname{chapter}{chapter}{chapters}
\crefname{section}{section}{sections}
\crefname{appendix}{appendix}{appendices}
\crefname{subsection}{section}{sections}
\crefname{subsubsection}{section}{sections}
\crefname{equation}{equation}{equations}
\crefname{figure}{figure}{figures}
\crefname{table}{table}{tables}
\crefname{subfigure}{figure}{figures}
\crefname{listing}{listing}{listings}

\newcommand{\sparkfall}{%
\begin{sparkline}{12}
\sparkrectangle 0.05 0.95
\spark 
   0.000   0.5 
   0.076   0.5
   0.077   0.9
   0.153   0.9
   0.154   0.5
   0.230   0.5
   0.231   0.1
   0.307   0.1
   0.308   0.5
   0.384   0.5
   0.385   0.775
   0.461   0.775
   0.462   0.5
   0.537   0.5
   0.538   0.225
   0.614   0.225
   0.615   0.5
   0.691   0.5
   0.692   0.65
   0.768   0.65
   0.769   0.5
   0.845   0.5
   0.846   0.35
   0.922   0.35
   0.923   0.5
   1.000   0.5 /
\end{sparkline}
}
\newcommand{\sparkrise}{%
\begin{sparkline}{12}
\sparkrectangle 0.05 0.95
\spark 
   0.000   0.5 
   0.076   0.5
   0.077   0.35 
   0.153   0.35
   0.154   0.5
   0.230   0.5
   0.231   0.65
   0.307   0.65
   0.308   0.5
   0.384   0.5
   0.385   0.225 
   0.461   0.225
   0.462   0.5
   0.537   0.5
   0.538   0.775
   0.614   0.775
   0.615   0.5
   0.691   0.5
   0.692   0.1
   0.768   0.1
   0.769   0.5
   0.845   0.5
   0.846   0.9
   0.922   0.9
   0.923   0.5
   1.000   0.5 /
\end{sparkline}
}

\begin{document}

\title{Influence of 
surfactants on the electrohydrodynamic stretching of water drops in
oil}
\date{\today}
\author[ept]{Åsmund Ervik\corref{cor1}}
\ead{asmunder@pvv.org}
\author[phys]{Torstein Eidsnes Penne}
\author[sintef]{Svein Magne Hellesø}
\author[sintef]{Svend~Tollak~Munkejord}
\author[ept]{Bernhard Müller}

\cortext[cor1]{Corresponding author}

\address[ept]{Department of Energy and Process Engineering, Norwegian University
of Science and Technology (NTNU), 7491 Trondheim, Norway}
\address[phys]{Department of Physics, Norwegian University
of Science and Technology (NTNU), 7491 Trondheim, Norway}
\address[sintef]{SINTEF Energy Research, P.O. Box 4761 Sluppen, 7465 Trondheim,
Norway}

\begin{abstract}

In this paper we present experimental and numerical studies of the
electrohydrodynamic stretching of a sub-millimetre-sized salt water drop,
immersed in oil with added non-ionic surfactant, and subjected to a suddenly
applied electric field of magnitude approaching 1 kV/mm. By varying the drop size,
electric field strength and surfactant concentration we cover the whole range of electric
capillary numbers ($Ca_E$) from 0 up to the limit of drop disintegration. The results
are compared with the analytical result by G.I. Taylor (\emph{Proc. R. Soc. A}
\textbf{280}, 383 (1964)) which predicts the asymptotic deformation as a function
of $Ca_E$. We find that the
addition of surfactant damps the transient oscillations and that the drops may be stretched
slightly beyond the stability limit found by Taylor. We proceed to
study the damping of the oscillations, and show that increasing the surfactant
concentration has a dual effect of first increasing the damping at low
concentrations, and then increasing the asymptotic deformation at higher
concentrations. We explain this by comparing the Marangoni forces and the
interfacial tension as the drops deform. Finally, we have observed
in the experiments a significant hysteresis
effect when drops in oil with large concentration of surfactant are subjected
to repeated deformations with increasing electric field strengths. This effect is not
attributable to the flow nor the interfacial surfactant transport.

\end{abstract}


\maketitle


\section{Introduction}
Surfactants are ubiquitous in two-phase fluid flows. Take for instance a single 
drop falling through a viscous fluid, perhaps the simplest and most widely
studied two-phase flow configuration. While the classic 
results by \citet{hadamard1911} and \citet{rybzynski1911} give the analytical
result for the flow field in this case, experimental investigations mostly fail
to agree with this result. The discrepancy is attributed to trace surface-active
contaminants, found even in the most purified of liquids. It is natural, then,
also to consider the effects of surfactants on the more complicated case of
electrohydrodynamic deformation of a conducting drop falling in an insulating
oil.

The case of a drop deforming in an electric field is interesting, not only as
an intriguing physical phenomenon of which our understanding can be improved, but also
for applications \eg to chemical processing equipment such as electrocoalescers
\citep{atten1993,eow2001,lundgaard2006}. A deeper understanding of the 
physical processes at play in this system could lead to improved coalescer 
equipment and reduced emissions.

We will consider here experiments and simulations of
sub-millimetre-sized drops of brine falling in a highly refined oil with added
surfactant, studying the drop deformations and oscillations induced by square
voltage pulses of varying amplitude applied to parallel electrodes above and
below such a drop.

When performing these studies of drop deformations, it is
crucial to have a system which is well characterised in terms of the fluid
and the interfacial properties. To overcome the uncertainties associated
with unknown trace contaminants acting as surface-active agents, we
deliberately add a non-ionic surfactant (Span 80) in known, small quantities. The
interfacial tension as a function of surfactant concentration is then
measured, together with the bulk properties, to give a well-characterised
system.

There is a large amount of research on the deformations of drops in electric
fields, using analytical, experimental and numerical techniques; we will not
summarise all of it here. The review by \citet{melcher1969} covers the
fundamentals in a thorough fashion, while the review by \citet{saville1997}
gives an update with more recent results in the field.  However, when
surfactants are added to this picture, the literature is not so extensive.
Previous authors \citep{ha1998,zhang2014a} have investigated the influence of
surfactants on the electrohydrodynamic stretching experimentally, but they have
been limited to considerations of the static (equilibrium) deformation, as well
as drop sizes above 1 mm in diameter, and a limited number of observations.
Computational studies in the literature, namely previous work by
\citet{teigen2010}, and the paper by \citet{nganguia2013} which finds good
agreement with \cite{teigen2010}, have also been focused on the static
deformation. Note that the numerical code used in this paper is the same as in
\cite{teigen2010}. 

Taking a step further, we consider here also the dynamical behaviour of the
stretching drops, in particular the effects of the surfactant concentration on
the damping of the drop oscillations. We work with drops smaller than 1 mm in
diameter. We report results for many drop deformations, almost 300 for the
experiments and 44 representative cases for the simulations.

This work is an extension of our initial investigation \citep{ervik2014b}, where
five cases of the electrohydrodynamic deformation of drops in insulating oil were
studied. In the present work we have extended this analysis to a parameter
study of the factors influencing the deformation and the deviations
from the classical result by \citet{taylor1964}, which does not take surfactants into
account. The analytical result by Taylor has been found to agree
very well with subsequent results, see \eg \cite{brazier1971}, and for this reason we use it as
a supporting line in the plots and analysis throughout the paper. Following
Taylor, we use the dimensionless electric field strength $\zeta = \sqrt{Ca_E}$
in the following. 

The results presented here show that the deviation from Taylor's expression is negligible below
dimensionless electric field strengths of $\zeta \approx 0.4$, while above this
threshold they become significant. We demonstrate that drops in the presence of
surfactants may be deformed beyond the stability limit given by the Taylor
theory. Finally we study the effect of the surfactant concentration,
and the effects of Marangoni stresses on the damping of drop oscillations. Our results indicate that small
concentrations of surfactant give a significant increase in the damping whilst
having but a small effect on the equilibrium (static) shape. Also, for the highest
surfactant concentration used here, we observe in the experiments a significant hysteresis effect
of repeated stretchings. This effect is not seen in the simulations, so it 
cannot be explained by the hydrodynamics and the surfactant transport processes
which are modelled by our approach.

\section{Theory}
\label{sec:theory}
The flow of single-phase oil or water can be described by the incompressible 
Navier-Stokes equations
\begin{align}
  \div\v{u} &= 0 \label{eq:ns-divfree} \textnormal{,} \\
	\pd{\v{u}}{t}+(\v{u}\cdot\grad)\v{u} &= - \frac{\grad{p}}{\rho}
  + \frac{\eta}{\rho}\grad^2\v{u} +
  \v{f} \textnormal{,}
	\label{eq:ns}
\end{align}
where $\v{u}$ is the velocity field, $p$ is the pressure, $\rho$ is the density,
$\eta$ is the dynamic viscosity, and $\v{f}$ is the acceleration caused by some
body force, \eg the gravitational acceleration.
This description can be extended to a two-phase flow by incorporating three
things, namely that there is an interface separating the two fluids, that
the fluids may have different viscosities $\eta_1,\eta_2$ and densities
$\rho_1,\rho_2$, and finally the
effects of interfacial tension and interfacial tension gradients. We mark
the drop properties with subscript $_1$ and the bulk properties with $_2$,
and denote the interfacial tension by $\gamma$. The viscosity difference
and the interfacial tension $\gamma$ contribute to jumps across the
interface in various properties such as the pressure; this is detailed in
\crefrange{eq:pjump}{eq:gradujump} below. Mathematically, this can be
incorporated into the Navier-Stokes equations as a singular contribution to
$\v{f}$ in \cref{eq:ns}.

This system admits two dimensionless groups, which we may take to be the Reynolds number
$Re$ and the Ohnesorge number $Oh$. The Reynolds number is of interest for a falling drop, where it is 
defined as $Re_{\text{D}} = \rho_2 u_T D / \eta_2$, $u_T$ being the terminal
velocity and $D$ being the drop diameter. For the drops
considered here, the Reynolds number is small ($Re_\text{D} < 1$), meaning that
the inertial term in \cref{eq:ns} is unimportant for the flow at terminal velocity.

For an oscillating drop, the Ohnesorge number is an important quantity; some
authors use the inverse of the Ohnesorge number as the ``oscillation Reynolds
number'' $Re_{\text{osc}}$. We use the definition $Oh
= \eta_2/\sqrt{\rho_2\gamma D}$, since the ambient fluid is much more viscous
for the cases considered here. For the oscillations, the Ohnesorge number is also small
($Oh < 0.2$), but here the inertial term is important since small $Oh$ 
corresponds to large $Re_\text{osc}$.

When considering a single small (\ie spherical) drop falling in a clean
fluid at low Reynolds number, the terminal velocity as well as the flow in
the entire domain is given analytically by the results that
\citet{hadamard1911} and \citet{rybzynski1911} obtained independently,
\begin{equation}
v_{\textnormal{T,HR}} = \frac{(\rho_1 - \rho_2) \v{g} D^2(\eta_1
+ \eta_2)}{6\eta_2(3\eta_1 + 2\eta_2)} \textnormal{.}
\label{eq:hadamard}
\end{equation}
Experimental results for the terminal velocity, however, tend to not agree with
this result \citep[see \eg][Fig. 1]{bond1928b}, but a closer agreement is found 
with the formula derived by \citet{stokes1851} for a hard sphere falling 
in an unbounded domain,
\begin{equation}
v_{\textnormal{T,S}} = \frac{(\rho_1 - \rho_2) \v{g} D^2}{18\eta_2} \textnormal{.}
\label{eq:stokes}
\end{equation}
Indeed Hadamard himself was aware of this discrepancy, as he states in his 1911
paper.

We note that for $\eta_1 < \infty$, the graphs of $v_{\textnormal{T}}(D)$ given
by \cref{eq:hadamard,eq:stokes} only intersect at $D=0$, and thus the terminal
velocity of a falling drop is an observable quantity that can determine if
a system is clean or not. An experimental observation closer to \cref{eq:stokes}
indicates a contaminated system, which is indeed the observation for most fluid
combinations. It is noteworthy that the experiments which have obtained values
agreeing with \cref{eq:hadamard} are for quite singular fluid combinations, \eg
mercury drops in glycerine \citep{levich1962}.

The currently accepted explanation \citep[see][pp.\ 35-41]{clift1978} of this
phenomenon is that trace contaminants in the system act as surfactants which are
swept along the interface by the flow, creating an interfacial-tension gradient
which results in a Marangoni force, with the end result that the drop interface
is immobile. Since the nature of these trace contaminants are not known, we
deliberately add to the oil a known amount of a non-ionic surfactant, Span 80,
such that we obtain a well-described fluid system. 

The interfacial tension, $\gamma$, can be related to the bulk concentration of
surfactant, $\Lambda$, using the \citet{szyszkowski1908} equation of state (EoS):
\begin{equation}
\gamma(\Lambda)=\gamma_0\left[1-\beta\ln\left(1+\frac{\Lambda}{a_L}\right)\right],
\label{eq:szyszkowski}
\end{equation}
where $\gamma_0$ is the interfacial tension without surfactants, $\beta
= R_{\textnormal{gas}} T \Gamma_\infty/\gamma_0$ is the interfacial elasticity, and
$a_L=k_{\textnormal{des}}/k_{\textnormal{ads}}$ is the ratio between the adsorption and
desorption coefficients of the surfactant. In the expression for $\beta$, $\Gamma_\infty$
is the maximum possible interfacial concentration of surfactant, $R_{\textnormal{gas}}$ is the universal
gas constant, and $T$ is the
temperature (in Kelvin). The parameters $\beta,a_L$ of this EoS may be computed by fitting
to experimental data; note that this also determines $\Gamma_\infty$ when the
temperature is known.

The equilibrium interfacial concentration can subsequently be calculated as
\begin{equation}
\Gamma = \Gamma_\infty\frac{\Lambda}{\Lambda+a_L}.  
\label{eq:Gamma}
\end{equation}
The relationship between interfacial concentration and interfacial tension
is then given by the Langmuir EoS:
\begin{equation}
\gamma(\Gamma)=\gamma_0\left[1
  +\beta\ln\left(1-\frac{\Gamma}{\Gamma_\infty}\right)\right].
\label{eq:langmuir}
\end{equation}
For a detailed review of these equations and their derivation, see \eg
\cite[pp. 47--50]{dukhin1995}. In the next section we plot
this equation, with parameters obtained by fitting \cref{eq:szyszkowski} to the experimental data of interfacial tension as
a function of concentration, together with the experimental data; it is seen
that the fit is very good.

In the present case we consider the surfactant to be insoluble, and we
restrict ourselves both in simulations and experiments to surfactant
concentrations which are below the critical micelle concentration (0.02
wt\% for our system). An insoluble surfactant is a good approximation
when the time scales for adsorption-desorption are long when compared
to the deformation time scales \citep{pawar1996,lucassen2001}. This
is the case here, since the time it takes to reach equilibrium for the
measurements of interfacial tension is in the order of minutes, while the time
period of the drop deformations discussed is on the order of milliseconds. We
denote the non-equilibrium interfacial surfactant concentration by $\xi$.
The initial value of $\xi$ is given by
\cref{eq:Gamma}, and the concentration profile $\xi(\v{x})$ evolves according to an
advection-diffusion equation which is restricted to the interface (see \eg
\cite{xu2006}), namely
\begin{equation}
  \pd{\xi}{t} + u_i\frac{\partial \xi}{\partial x_i} - n_i n_j
  \frac{\partial u_j}{\partial x_i}\xi =
  D_{\xi}\left(\frac{\partial^2\xi}{\partial x_i\partial x_i}
    - \frac{n_i n_j \partial^2\xi}{\partial x_i\partial x_j}
  + \kappa\,n_i\frac{\partial \xi}{\partial x_i}\right),
\label{eq:advec-diff}
\end{equation}
where we employ the Einstein summation convention. $u_i$ and $n_i$ denotes the
components of the velocity $\v{u}$ and the normal vector $\v{n}$, respectively. $\kappa = \div \v{n}$ is the
interfacial curvature. $D_{\xi}$ is the surfactant
interfacial diffusion coefficient, a parameter which is very
difficult to measure. Fortunately the solutions of this equation are
quite insensitive to the value of this constant as long as it is small. Here we use the value
$5\cdot10^{-7}$ m$^2$/s, which is of the same magnitude as that reported \eg by 
\citet{sakata1969} (albeit for different surfactants). 

With this approach, the Gibbs elasticity is taken into account, and its
magnitude can be computed as $\beta\gamma$ \citep{lucassen2001}. This
only takes into account the elasticity caused by the change in interfacial
tension given by a change in the drop area. Other physical mechanisms, such as
reorientation of surfactant molecules at the interface, can lead to
additional effects, and even cause a phase transition in the surfactant layer
\citep{ravera2005}.

A seminal approach to the stretching of drops by electric fields is the
study by \citet{taylor1964} who did a theoretical analysis of the
electrohydrodynamic stretching of a clean conducting drop in a perfect
dielectric medium. His result predicts the asymptotic drop
deformation as
a function of the electric field strength, radius, permittivity of the oil and
interfacial tension, all combined into a dimensionless electric field strength
$\zeta$. This is equivalent to the square root of the electric capillary number,
$\zeta = \sqrt{Ca_{E}}$, where $\zeta$ is defined as 
\begin{equation}
  \zeta = \bar{E}\sqrt{\epsilon\epsilon_0 D/\gamma} \textnormal{,}
\label{eq:taylor}
\end{equation}
and $\bar{E}$ is the uniform electric field that is present far away from the
drop. Note that papers from that era work in electrostatic units, where the
numerical value of $\epsilon_0$ is 1, so it is frequently omitted from their
formulae. Note also that some authors use the drop radius rather than the
diameter here.

We may compare $\zeta^2$ to the capillary number computed from the terminal
velocity, $Ca = \eta_2 u_T / \gamma$, giving us an impression of the relative
importance of the external flow versus the electric field as far as the drop
shape is concerned. Using numbers relevant to the
situation at hand, we estimate a typical value of the hydrodynamic capillary number
to be $Ca \approx 0.007$, while a typical value for the electric
capillary number is $\zeta^2 \approx 0.25$, indicating that the electric field has a much
greater influence on the drop shape than the deformation due to the external
flow. One may thus neglect the effects of the external flow when considering the
drop deformations.

In his analysis of drop deformation, Taylor considered a clean drop at rest in
a medium with the permittivity of free space, approximated the drop shape at
equilibrium as ellipsoidal, and assumed that the value of the difference between
interfacial tension and electrical stress at the interface is equal at the poles
and at the equator. He then derived an implicit formula which predicts the drop
elongation $a/b$ as a function of $\zeta$. In this context, $a$ and $b$ denote
the semi-major and semi-minor axis of the drop, respectively, see also
\cref{fig:frame_180} in the next section. 

\begin{figure}
  \begin{center}
    \includegraphics[width=0.7\linewidth]{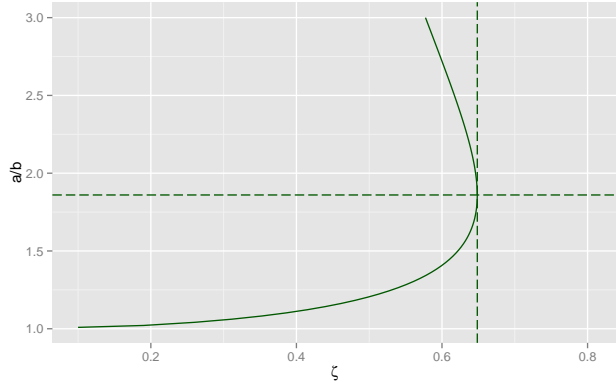}
  \end{center}
  \caption{The static deformation predicted by Taylor's theory.}
  \label{fig:taylor}
\end{figure}

That implicit
formula may be given \eg as the zero level of the function
\begin{align}
  f\left(\zeta,\frac{a}{b}\right) &=
  2 \left(\frac{a}{b}\right)^{-4/3}\sqrt{2
  - \left(\frac{a}{b}\right)^{-1} - \left(\frac{a}{b}\right)^{-3}} I
 - \zeta \textnormal{,} \\
 I &=  \frac{1}{2}e^{-3}\ln\left( \frac{1+e}{1-e} \right) - e^{-2} \textnormal{,} \\
 e &= \sqrt{1 - (a/b)^{-2}} \textnormal{.}
\end{align}
This predicts a
limit to the static deformation at $a/b \approx 1.86$ and $\zeta \approx
0.65$; at higher applied field strengths the drop does not reach an
equilibrium state, but is torn apart. Taylor showed that this limit agrees
with experiments done with drops in air. The theoretical result by Taylor is shown in
\cref{fig:taylor} together with a horizontal line at $a/b = 1.86$ and a
vertical line at $\zeta = 0.6485$.

For the numerical model, including the effects of an electric field on the drop
requires knowledge of this field inside the simulation domain. Even though we
consider a conducting drop in a dielectric medium, we may model the situation as
two dielectric media with a very high permittivity ratio
\citep{melcher1969}. We use here a numerical value of 1000 for the relative
permittivity of the conducting liquid; this value is not important as long as it
is much larger than that of the dielectric liquid. The model validity is
confirmed by the fact that the calculated field lines inside the drop are
indeed straight, parallel lines in the direction normal to the
electrodes, as seen in \cref{fig:sim-oscillations-b}.

To obtain the electric field, we may then proceed by solving a Laplace equation
for the electric
potential $\Psi$, with the applied voltage as boundary conditions at the top
and bottom of the domain, and $\grad \Psi \cdot \v{n} = 0$ at the vertical
boundaries of the domain. To wit:
\begin{equation}
\div(\epsilon\epsilon_0\grad\Psi) = 0,
\label{eq:laplace}
\end{equation}
where $\epsilon$ is the relative permittivity and $\epsilon_0$ is the
permittivity in vacuum. Note that we keep $\epsilon$ inside the
divergence operator here, even though it is piecewise constant, since this is
how the discontinuity is handled by the numerical method.

The Maxwell stress tensor, $\tensor{M}$, can then be
calculated from the electric field $\v{E} = - \grad \Psi$. Neglecting
the magnetic field, which is not of interest here, we have
\begin{equation}
\tensor{M}
= \epsilon\epsilon_0\Bigl(\v{E}\v{E}-\frac{1}{2}(\v{E}\cdot\v{E})\tensor{\v{I}}\Bigr) \textnormal{.}
\label{eq:Mtensor}
\end{equation}
This stress gives a spatially varying
contribution to \eg the jump in the pressure across the interface, which will distort
a drop from its spherical shape.

All in all, the formulation presented here takes into
account the effects of interfacial tension $\gamma$, the applied electric
field, and the Marangoni effect that arises from an interfacial-tension
gradient. The jumps across the drop interface in various properties are then
given as \citep{bjoerklund2008,teigen2010}
\begin{align}
\llbracket \v{u} \rrbracket &=0 \textnormal{,}\\
\label{eq:pjump}
\llbracket p \rrbracket &=2\llbracket \eta \rrbracket \v{n}\cdot
{\grad \v{u}}\cdot\v{n} + \v{n}\cdot \llbracket
\tensor{M} \rrbracket \cdot \v{n} - \gamma\kappa \textnormal{,}\\
\llbracket \Psi \rrbracket  &= 0 \textnormal{,}\\
\label{eq:gradujump}
\llbracket \eta {\grad \v{u}}\rrbracket &=\llbracket \eta \rrbracket \Bigl(
(\v{n}\cdot {\grad \v{u}}\cdot\v{n})\v{n}\v{n}+(\v{n}\cdot {\grad
\v{u}}\cdot \v{t})\v{n}\v{t} \nonumber \\
&\qquad\;- (\v{n}\cdot {\grad \v{u}}\cdot \v{t})\v{t}\v{n}+(\v{t}\cdot
  {\grad \v{u}}\cdot \v{t})\v{t}\v{t} \Bigr) \\
&\quad - (\v{t}\cdot
\grad_\iota \gamma)\v{t}\v{n} \textnormal{,} \nonumber
\end{align}
In these expressions, $\v{n}$ and $\v{t}$ are the normal and tangent unit
vectors to the interface. Expressions such as ${\grad \v{u}}$ and
$\v{n}\v{n}$ denote rank-two tensors formed by the dyadic product, so
\eg ${\grad \v{u}}\cdot\v{n}$ denotes such a tensor acting on a vector.  We use the
convention that a normal vector on a drop points towards the external
fluid, and that the jump $\llbracket-\rrbracket$ is the difference
between the external and the internal properties, \eg $\llbracket \eta
\rrbracket = \eta_2 - \eta_1$. 
The interface is denoted by $\iota$ here,
so $\grad_\iota$ is the gradient along the interface. For the sake of
completeness, we mention that an
additional term $- (\v{t}\cdot \llbracket \tensor{M}\rrbracket\cdot \v{n})$
contributes to \cref{eq:gradujump} when the fluids are not perfect
dielectrics, but rather leaky dielectrics. This is considered \eg by
\cite{teigen2010a,sunder2016}, and gives an electric contribution to the
tangential force at the interface.

\section{Methods}
\label{sec:methods}
\subsection{Experimental methods}
\label{sec:expmeth}
Experiments were performed with brine drops (3.36 wt\% NaCl added to Milli-Q
purified water) immersed in Marcol 52 oil (ExxonMobil), which is a purified and hydrogenated
hydrocarbon oil with very low content of surface-active components. Span 80
non-ionic surfactant (Sigma-Aldrich) was added. The densities were measured with an Anton Paar DMA 5000
density meter. The viscosity of the oil was measured with an Anton Paar MCR 102
rheometer. Tabulated values from \citet{white} were used for brine
viscosity. Experiments were done with temperature control at
21.5\textdegree C, where viscosity and density of water were 1.03
mPa$\cdot$s and 1023.6 kg/m$^{3}$, respectively, and those of the oil were
12.4 mPa$\cdot$s and 832.3 kg/m$^{3}$, respectively.
The relative
permittivity of Marcol 52 was taken to be 2.13, as per the data sheet
supplied by the manufacturer.

Interfacial tension
was measured with a SIGMA 703D tensiometer with a DuNuoy ring, for
different Span 80-concentrations, with selected values shown in
\cref{tab:IFT}. All data points are given in the supplementary information. Here wt\% means weight percent. From these measurements,
the critical micelle concentration was determined to be 0.020 wt\%, and we
limit ourselves to concentrations below this value. 
\begin{table}
  \centering\footnotesize
\begin{tabular}{cc}
  \toprule
  wt\% Span 80 & Interfacial tension [mN/m] \\ 
\midrule
0.030 & 10.0 \\ 
0.020 & 10.1 \\ 
 0.015 & 13.9 \\ 
0.010 & 18.8 \\ 
0.001 & 29.4 \\ 
\bottomrule
\end{tabular} 
\caption{Interfacial tension between water and oil for different surfactant concentrations}
  \label{tab:IFT}
\end{table}

\begin{figure}
  \begin{center}
    \includegraphics[width=0.8\linewidth]{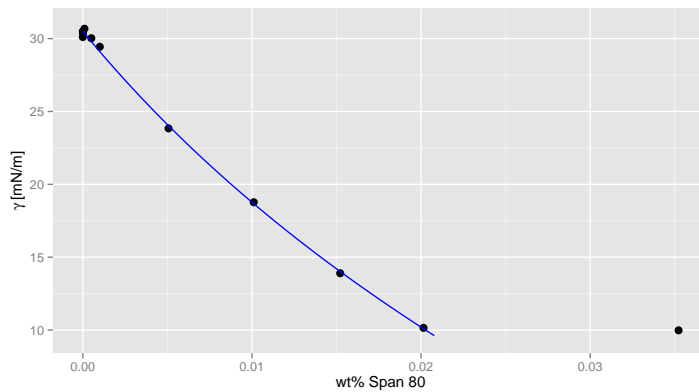}
  \end{center}
  \caption{Experimental measurements of interfacial tension (points) and the
  Langmuir EoS~(\ref{eq:langmuir}) fitted to these (line). Note that the
interfacial tension is constant above 0.02 wt\%, indicating that this is the CMC.}
  \label{fig:langmuir}
\end{figure}

$\beta$ and $a_L$ in \cref{eq:szyszkowski} were determined by fitting
this equation to the experimental measurements
using non-linear least-squares. See the plot of the data points and the fitted
equation in \cref{fig:langmuir}; note in particular that the interfacial tension
is constant above 0.02 wt\%, confirming that this is the critical micelle
concentration (CMC). It is somewhat
difficult to tell whether the point at 0.02 wt\% is a little above or a little below the
CMC; thus we have tested the sensitivity of the curve fit to this point by also
computing a fit with this point omitted. At the highest concentration studied
here, 0.016 wt\%, this change in the curve fit produced a change in the
interfacial tension predicted by the EoS of 0.1 mN/m, \ie less than 1\% and
within the experimental uncertainty. Accordincly, the small uncertainty about the exact
value of the CMC has no influence on the results presented in this paper.

In addition to fitting the EoS, the interfacial area available to
each surfactant molecule at the critical micelle concentration, $A_{\textnormal{CMC}}$, was estimated
from the slope of the Gibbs isotherm as it approaches the CMC \citep{tsujii1998}. For further details, see the supplementary
information which contains the script used for fitting and the
experimental data. When comparing with the results by \citet{peltonen2000}, we
find good agreement for the values of $A_{\textnormal{CMC}}$ and the critical micelle
concentration (CMC) obtained here, $30.5$ Å$^2$ and 0.020
wt\%, respectively.

The deformation of the water drops was observed as they fell in the 15 mm gap between an
upper and a lower horizontal metal electrode. Drops were produced from
a screw-in syringe connected to a glass
capillary tube made hydrophobic by a silane coating; this tube protruded through
a small hole in the upper electrode. A series of square voltages with
different amplitudes was applied to the lower electrode, creating
an electric field $\v{E}$ that distorted the drop. The voltage pulse shape was
generated in MATLAB and sent over a serial connection to a Stanford Research
DS340 signal generator connected to a TREK 2020B high voltage amplifier. The
voltage pulse shapes were either rising or falling, and included both positive
and negative pulses; \ie the pulses were bipolar. The length of each pulse was 25 ms, with a 25 ms pause
between pulses. The amplitudes were defined as fractions of a maximum
amplitude $V_0$, \eg $V_0 =$ 10 kV and fractions 2/4, 3/4 and 4/4 times $V_0$,
giving pulses that look like \sparkrise or \sparkfall. Typically
6 different amplitudes (fractions) were used.
 This
application of different voltages resulting in different stretchings of the
same drop is the only practical way of studying the effect of varying the
electric field strength at constant drop radius.
It also allows us to study possible hysteresis effects of the stretching on
the surfactant on the interface. Such effects have been reported previously,
\eg by \cite{peltonen2000} for the Span 80 surfactant used here.

When applying several voltage pulses to the electrodes it is desirable
to keep the drop in the camera field-of-view for as long as possible. To
achieve this, a moving stage setup was used, comprised of a Newport (M-)IMS-V
linear stage to move the test  cell containing the fluid system upwards a constant
velocity, and a Newport XPS Series Motion Controller to manually match the velocity of
the moving stage to the terminal velocity of the drop.

\begin{figure}
   \centering
   \begin{subfigure}[b]{0.3\linewidth}
   \includegraphics[width=\linewidth]{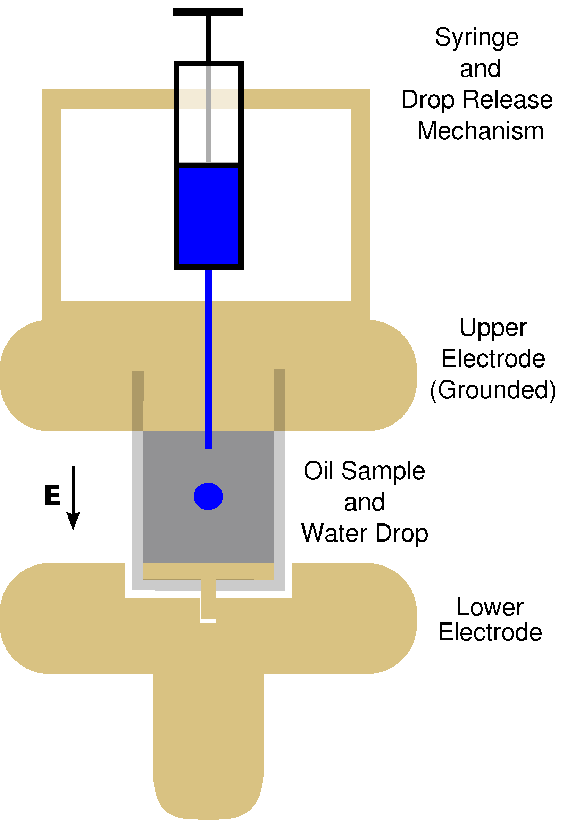}
   \caption{Experimental setup.}
   \label{fig:test-cell}
   \end{subfigure}
   $\;$
   \begin{subfigure}[b]{0.3\linewidth}
    \includegraphics[width=\linewidth]{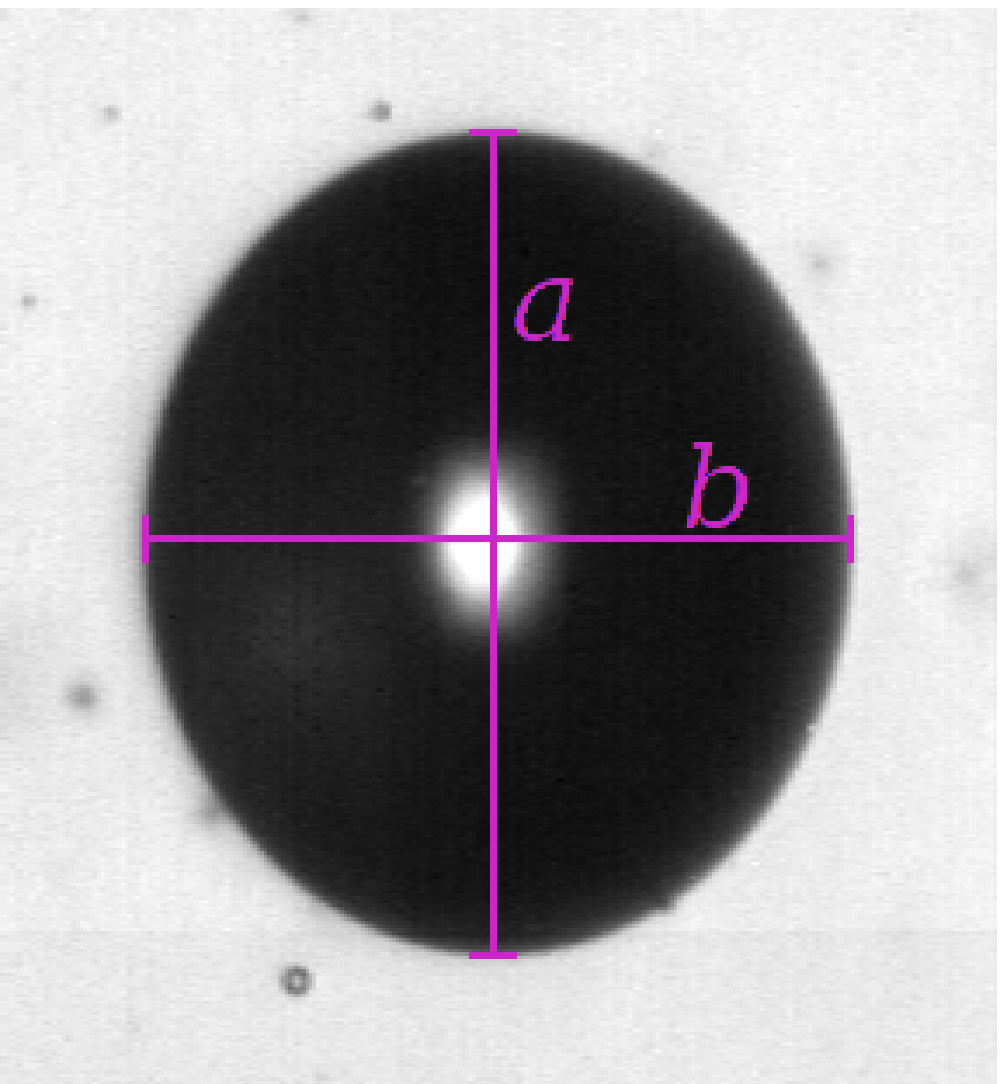}
    \caption{Captured image of drop.}
    \label{fig:frame_180}
   \end{subfigure}
   $\;$
   \begin{subfigure}[b]{0.3\linewidth}
    \includegraphics[width=\linewidth]{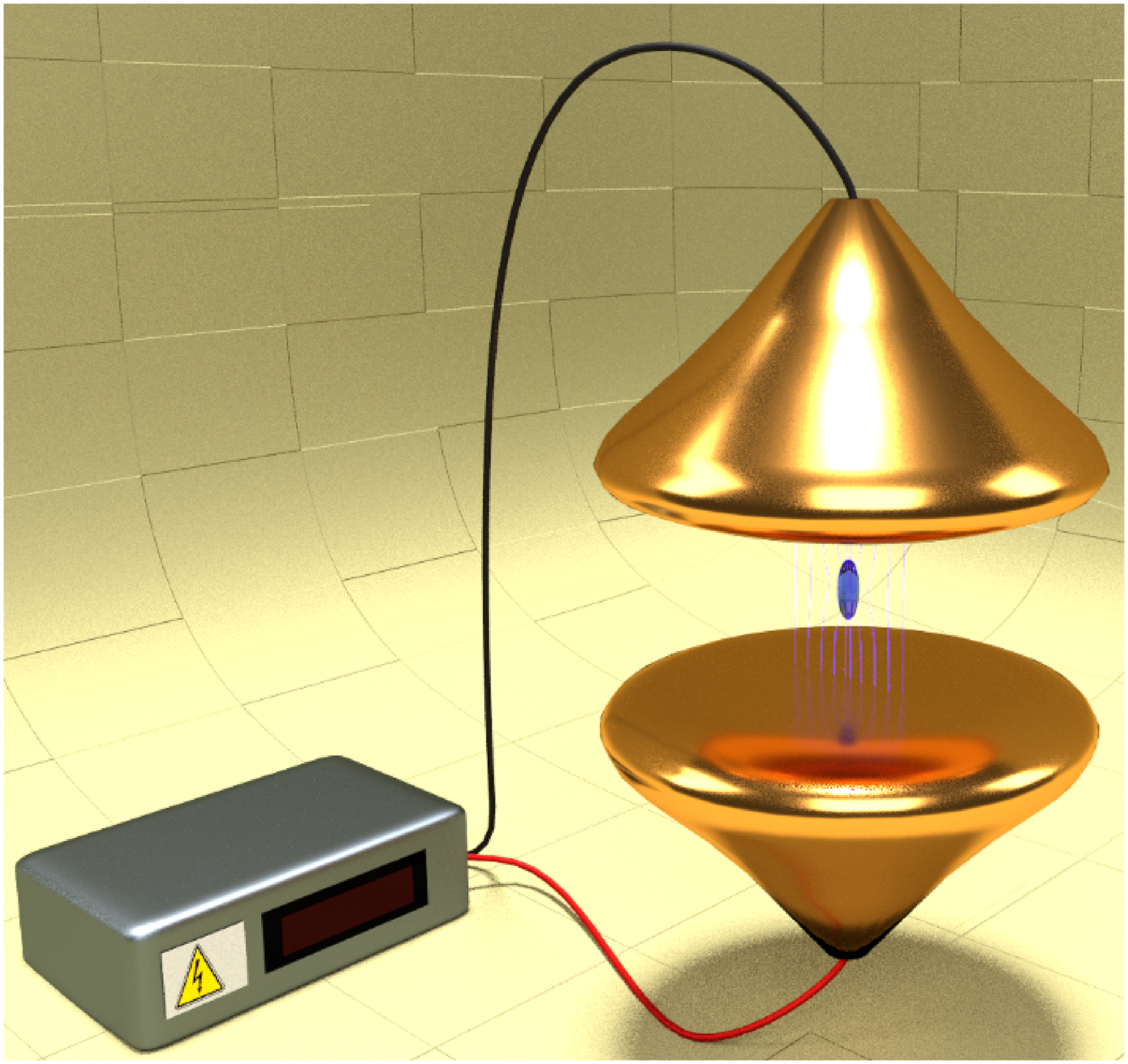}
    \caption{3D rendering of setup.}
    \label{fig:test-cell-3D}
   \end{subfigure}
   \caption{%
      Schematic showing a side view of the experimental setup, an
      example of a captured drop image shown with the major and minor axes $a$
      and $b$ superimposed, and a 3D rendering with a simulated droplet and
      electric field shown between the electrodes. The drop size is
      exaggerated in both (a) and (c).
    } 
    \label{fig:exp-setup}
  \end{figure}

A side view of the experimental setup is shown in 
\cref{fig:test-cell}, and a 3D rendering is shown in
\cref{fig:test-cell-3D}. To avoid unnecessary clutter, the temperature control
bath, optical setup and the linear stage are omitted in both of these figures, and in \cref{fig:test-cell-3D} the cuvette containing oil and the syringe mechanism are
also omitted. The drop size relative to the setup is exaggerated in both
figures.

To record high-speed movies of a falling drop, a Cheetah CL near-infrared
camera was used with an Infinity KS2 long-range microscope lens, and
a collimated light
source was placed on the opposite side of the cuvette. The camera
recorded a frame of $640 \times 512$ pixels at 1730 frames per second.
The high-speed movies were recorded in the
Streams 7 software, together with the voltage pulse from the signal generator
and the velocity and position of the moving stage.
These electrical signals were captured using a National Instruments PCI-6052E DAQ board.

To determine the drop deformations from the high-speed
images, the Spotlight image analysis software was used.
With this software, the
dimensions of the major axis $a$ and minor axis $b$ of a deformed
drop can be determined, as seen in \cref{fig:frame_180}, where the axes are superimposed on an
image of an elongated drop. This then gives the ratio $a/b$ as a measure
of the deformation; note that this measure does not make any assumptions about
the drop shape. See also the supplemental material in movie 1 which shows a video
of a drop deformation cycle together with an animated plot of $a/b$ as
a function of time.

The various uncertainties that affected the measurement of $a/b$ were
analysed in a fashion similar to that used by \cite{zhao2009phd} and
Gaussian error propagation was then used to compute the uncertainty in
$a/b$ \citep{moffat1988}. This uncertainty was found to be independent of
$a/b$, but dependent on the initial drop radius, which is sensible. The
relative error in $a/b$ was largest for the smallest drops under
consideration, at 3.4\%, and smallest for the largest drops considered, at
2.0\%.
 
The experimental procedure for studying the drop deformation in an electric
field was comprised of the following steps:\\
\textbf{1.} Move the test cell using the linear stage such that the 
tip of the needle used for generating droplets is in the top of the camera's field-of-view. 
Use the screw-in plunger to create a droplet of the desired size.\\
\textbf{2.} Wait for 1 minute to allow the equilibration of surfactants at the drop
interface. Arm the camera such that it starts recording 10 ms before the first
voltage pulse is applied.\\
\textbf{3.} Use an electromagnet to jerk the glass needle upwards, releasing the
drop.\\
\textbf{4.} As the drop falls through the view of the camera, adjust the upwards
velocity of the moving stage to match the terminal velocity of the drop, keeping
the drop in the centre of the image. The
drop falls for approx. 0.5 seconds before the voltage pulse train is applied, allowing ample time for any initial oscillations to be damped away.\\
\textbf{5.} Trigger the voltage pulse train. The camera is also controlled by this 
trigger and records a movie of the drop being deformed.\\ 
\textbf{6.} Post-process the recorded movie to extract $a$, $b$ as functions of
time.\\

A remark is in order with regards to the waiting time for equilibration
of surfactants at the drop. As stated, a waiting time of one minute is used in
these studies. If we are to compare this with some intrinsic time scale, we may
consider $\tau_D = \Gamma^2/(\Lambda^2 D_B)$ \citep{lin1990}, with $\Gamma$ taken
at some surfactant concentration $\Lambda$,
say the highest used in these experiments (0.016 wt\%). The value of $D_B$, the
bulk diffusion coefficient of Span 80 in Marcol 52 oil, is not known. Since the
Span 80 molecule is not much larger than the alkanes in the oil, we may use as
a rough estimate the self-diffusion coefficient of the tail of the Span 80
molecule, namely oleic acid, which gives $D_B \approx 10^{-10}$ m$^2/$s
\citep{iwahashi2007}; of the same order of magnitude as \eg the diffusion coefficient of
C$_{12}$E$_6$ surfactant in water \citep{lin2003}. This gives a time scale of $\tau_D \approx 3 \cdot 10^5$ s,
\ie 3.5 days. It would be impractical to wait for such
a long time between each drop was produced.

Fortunately, the transport of surfactants to the interface is greatly
accelerated once the drop starts falling, since the velocity boundary layer
decreases the length of the diffusion boundary layer. Note that the length of
the diffusion boundary layer as estimated above is $l = \Gamma/\Lambda = 20
\mu$m, which is comparable to the drop radius of 250--500 $\mu$m.  To quantify
the increase in surfactant transport to the interface, we may consider the
Schmidt number, \ie the ratio between the viscous and the molecular diffusion
rates, which in this system is $Sc = 2 \cdot 10^5$. This indicates that the
combination of the one minute waiting time and the subsequent falling time of
0.5 s between drop release and the start of the first deformation should be
sufficient to ensure the interfacial surfactant concentration is equilibrated
before the deformations commence.

\subsection{Simulation and numerical methods}
\label{sec:simmeth}
An in-house code was used to solve the Navier-Stokes equations
(\ref{eq:ns-divfree}) and (\ref{eq:ns}) numerically. The simulations
reported here are done in axisymmetry. See 
\citet{teigen2009,teigen2010,teigen2010a} for validation of the methods and
implementation used here.

The equations are discretised on a structured, uniform, staggered grid
using the finite-difference method. For the convective terms, the fifth-order WENO
scheme \citep{jiang2000} is employed. For the other terms a standard second-order
central-difference scheme is used. The pressure and velocity fields are
coupled using the classical projection method due to \citet{chorin1968},
which gives a Poisson equation for the pressure. This Poisson equation is
solved here using the BoomerAMG (Algebraic MultiGrid) preconditioner \citep{henson2000} and the
BiCGStab (Bi-Conjugate Gradient Stabilised) iterative solver \citep{vandervorst1992}; we use the Hypre
\citep{hypre} and PETSc \citep{balay1997} libraries for these methods.

The equations are then
integrated in time using an explicit Runge-Kutta method which has the
strong stability preserving (SSP) property, namely SSPRK(2,2) in the terminology
of \citet{gottlieb2009}. Although this method is second order in time, the
overall scheme is only first order in time due to the irreducible splitting
error from the Chorin projection method.
To sum up, the present method is
first-order in time and second-order in space. 

To capture the position of the interface between the two fluids, the
level-set method \citep{osher2001} is used with the
high-order constrained reinitialisation method \citep{hartmann10}
and the 
velocity extrapolation procedure \citep{adalsteinsson1999}. With the interface
position known, the ghost-fluid method \citep{fedkiw1999} is used to enforce the jumps
specified in \crefrange{eq:pjump}{eq:gradujump} across the interface in
sharp fashion.

This formulation takes into
account the effects of interfacial tension $\gamma$, the applied electric
field, and the Marangoni effect that arises from an interfacial tension
gradient. The surfactant concentration along the
interface, $\xi$, is determined by solving the advection-diffusion 
\cref{eq:advec-diff}. The interfacial tension $\gamma$ is then determined by
the surfactant concentration according to \cref{eq:langmuir}, using
\cref{eq:Gamma} with $\xi$ in place of $\Lambda$.


\section{Results}
\label{sec:results}

\subsection{Parameter studies on drop deformation}
\label{sec:pardesign}
As stated previously, five distinct cases of drop deformation in the
presence of surfactants were studied in our previous paper \citep{ervik2014b},
along with studies on the terminal velocity. Detailed comparisons between
the experimental results and the simulation results are given there. In light of
the close agreement found, an experimental
and a computational parameter study were set up to better understand the effects
of surfactants. The aim of the present study is to give a sufficient coverage of
the parameter space, and to leverage the combination of simulations and
experiments to give a deeper insight. In particular
we study the initial drop oscillations when the pulse is applied, and the effect
the Marangoni force has on the damping of these oscillations, which has not been
studied before. The dataset generated by this study
is available in the supplementary information.
 
\begin{table*}
  \centering\footnotesize
\begin{tabular}{cccrc}
  \toprule
&Diameter ($\mu$m) & Electric field (V/mm) & Span 80 concentr. (wt\%) & $\zeta$
(-) \\ 
\midrule
Sim. & $(500,700,900)$ & $(300,500,700,900)$ & $(0.0,0.001,0.005,0.016)$
& $[0.15-0.80]$ \\
Exp. & $[578-902]$ & $[207-747]$ & $(0.001,0.005,0.016)$ & $[0.15-0.65]$ \\
\bottomrule
\end{tabular} 
  \caption{Ranges for parameters used.}
  \label{tab:params}
\end{table*}

For the simulations, a combination of three drop radii, four electric field strengths and four
surfactant concentrations was chosen, representative of the parameter ranges
used in the experiments. The dimensionless field strength $\zeta$ was then
computed for each combination, and additional combinations were added to
ensure a good coverage of the $\zeta$ values. The values are summarised in \cref{tab:params}.

From the experimental point of view, the surfactant concentration is also
well-defined at three values (the clean system cannot be reached). The drop
radius, on the other hand, is a quantity most difficult to control from one
drop to the next, so there is
no systematic variation in it. Finally the applied electric field strength is
defined from a base value and several fractions of this value, \eg
$(2/6,3/6,4/6,5/6,6/6)$. This is thus more controlled than the radii. But the
base value was varied for different interfacial-tension values, due to a desire
to avoid stretching drops beyond their stability limit, as drop destruction
necessitates stopping the experimental campaign and cleaning the test cell. The
values used in experiments are also summarised in \cref{tab:params}.

It should be noted that the simulations are all done
independently, while the experiments are done with several applications
of fields of different strength
on the same drop. Thus the simulations neglect any
hysteresis effects that arise from the hydrodynamics, \eg
if the flow caused by the previous deformation is still significant when the next
one commences. Also, since the scaling analysis of the capillary numbers
presented in \cref{sec:theory} indicates that gravity (\ie the external flow due
to falling) is unimportant for the
deformations, the simulations are performed with zero gravity. Both of these
assumptions are confirmed by simulating a falling drop subjected to a rising
voltage pulse, showing that neither the simplification of
zero-gravity nor that of independent deformations has a significant
effect on the simulation results.

\begin{figure}
  \begin{center}
   \begin{subfigure}[t]{0.45\linewidth}
    \includegraphics[width=\linewidth]{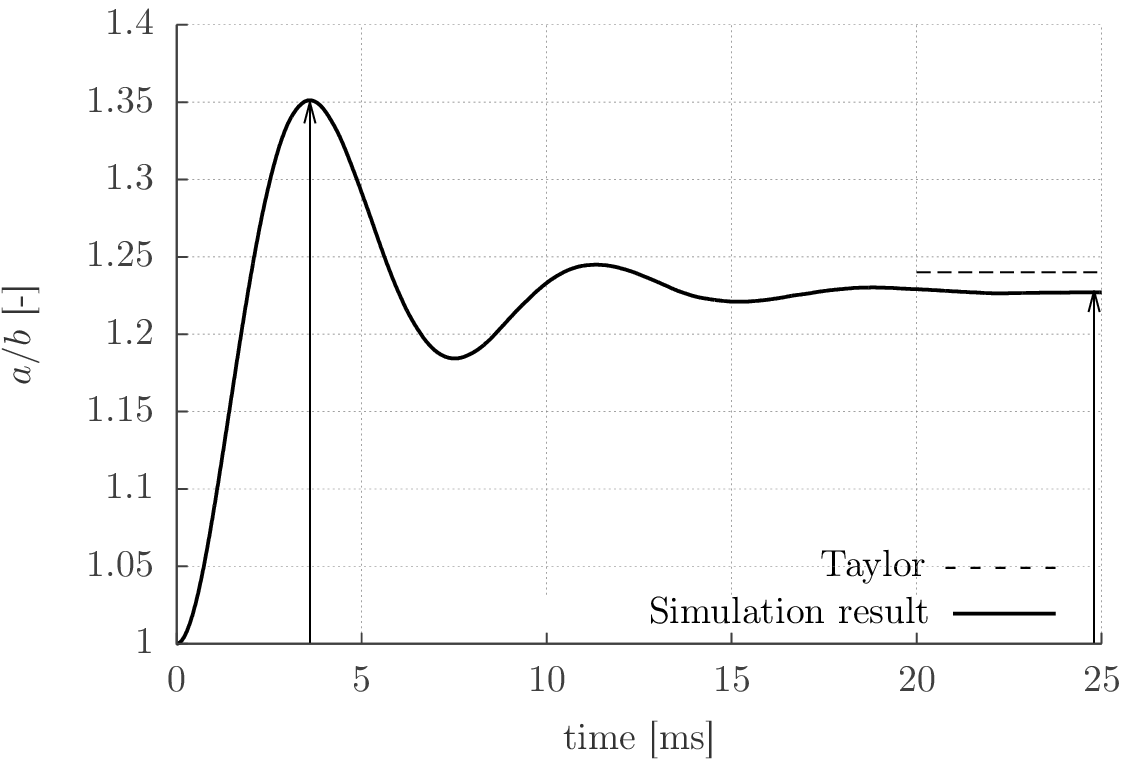}
    \caption{The deformation, $a/b$, plotted as a function of time. The peak and
      static values are pointed out by arrows, and the result by Taylor for the static
      deformation is shown as a dotted line on the right.}
    \label{fig:sim-oscillations-a}
  \end{subfigure}\hfill
   \begin{subfigure}[t]{0.45\linewidth}
    \includegraphics[width=\linewidth]{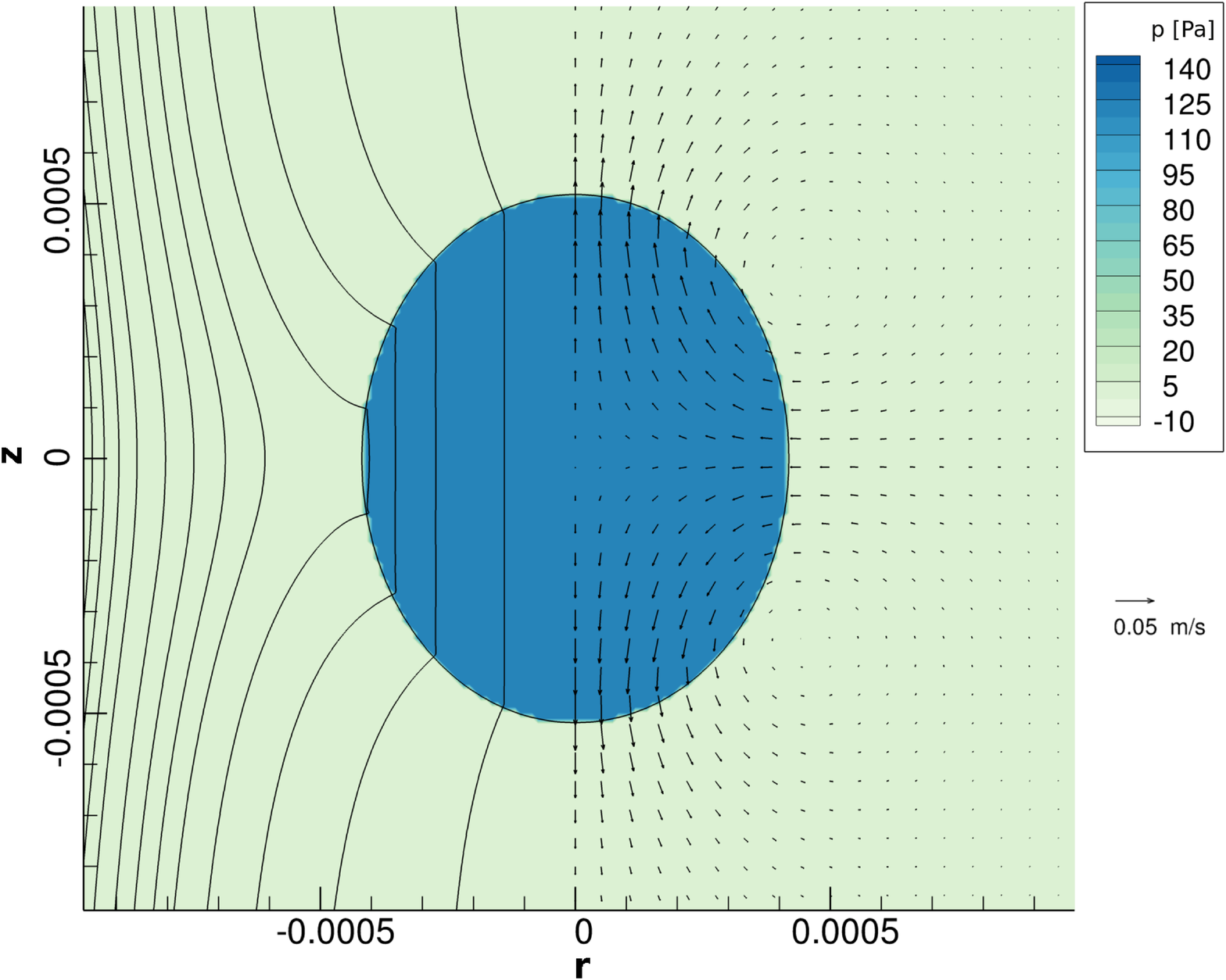}
    \caption{Snapshot of the deforming drop after 2 ms, halfway to the peak
      deformation. On the right side, velocity vectors are shown for every fifth grid
      point. On the left side, electric field lines are
      shown, quadratically spaced due to axisymmetry. The colour indicates pressure.}
    \label{fig:sim-oscillations-b}
  \end{subfigure}
  \end{center}
  \caption{Example case for $\zeta = 0.52$, a 0.9
mm diameter drop is being deformed by a 700 V/mm field without any surfactants
present.}
  \label{fig:sim-oscillations}
\end{figure}

In the subsequent sections we show plots of the deformation $a/b$, as discussed
in \cref{sec:theory}. During the deformation of a drop, $a/b$ starts as 1, then
increases to a peak value, and finally settles at some static value after some
oscillations. In \cref{fig:sim-oscillations} we plot the time evolution of $a/b$
for one of the cases considered here, as an example. We also show a snapshot of
the pressure, velocity and electric fields after 2 ms, which is halfway to the
peak deformation. Note that the maximum time in this line plot is the same as the
duration of a pulse, so this plot indicates the relaxation towards the new
equilibrium of a stretched drop.  See also the accompanying movie 1, where
a plot like \cref{fig:sim-oscillations-a} from one experiment is shown
side-by-side with the high-speed footage of a drop.

\subsection{Computational parameter study}
\label{sec:parsim}
In this section we report the results of parameter studies of the deformation
as a function of the dimensionless electric field strength. In total 44 cases
were studied. The simulations were performed in axisymmetry, using
a 241\x482 grid covering a $3D \times 6D$ domain. This is about six times larger in
each direction than what is shown in \cref{fig:sim-oscillations-b}. 
As discussed previously we
perform the simulations with zero gravity. The initial condition is then
a circular drop at rest, with an initial surfactant concentration given by the
bulk concentration according to \cref{eq:Gamma}.  The electric field is switched
on at $t=0$; the time it takes in experiments for the voltage to reach its
constant value is much smaller than the $\sim
0.5$ ms it takes for the drop to start deforming.  See also the supplemental
material in movie 2, where we show an animated 3D rendering produced from one of
these simulations, namely of the 900 $\mu$m diameter drop in 0.005 wt\% Span 80
subjected to a 700 V/mm electric field, corresponding to $\zeta = 0.58$.

\begin{figure}
  \begin{center}
    \includegraphics[width=0.8\linewidth]{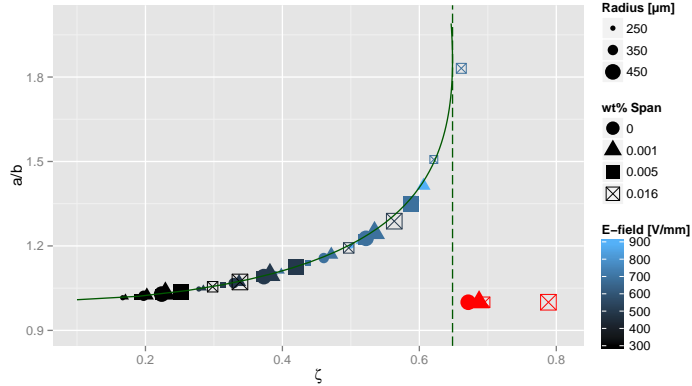}
  \end{center}
  \caption{The deformation $a/b$ found in simulations as a function of the
  dimensionless electric field strength, $\zeta$. Solid line: Taylor theory.
  Dashed line: static deformation limit. Red points: drop breakup.}
  \label{fig:sim-taylor}
\end{figure}

The results for the static deformation obtained in the simulations are shown in
\cref{fig:sim-taylor}. Here the static deformation for each case is shown as
a point and compared with the line which is the Taylor result. The points are
colour coded by the electric field. The shape of the points indicates the surfactant
concentration, and the size of the points indicates the drop radius. The points shown
in red were unstable, \ie the drop stretched until breakup.

As is seen from \cref{fig:sim-taylor}, there are deviations from the
Taylor theory, occurring mostly for large values of the dimensionless field strength. It is also seen
that the dimensionless parameter $\zeta$ is still a good variable for describing
the system in the presence of surfactants. We note that previous simulations
by \citet{brazier1971} also find some slight disagreement with the results by Taylor. This is
discussed further in \cref{sec:discussion}.

To further study the effects of increasing surfactant
concentration on the oscillation of drops, one may consider the analogy to
a damped mass-spring system. In that case, it is more convenient to work with
$a-b$ rather than $a/b$. This is because the former is directly related to the magnitude
$c_2$ of the fundamental mode of the oscillation, given by the coefficient of
the  second spherical harmonic, \emph{viz.} $a-b = \sqrt{45/16\pi}\times c_2$. Here we assume that
essentially only the second spherical harmonic contributes to the oscillation.
Note that under the typical assumptions in analytical work on drop oscillations,
as used \eg by \citet[pp. 473-475]{lamb1945}, the temporal evolution of $c_2$ corresponds
exactly to the evolution of a damped harmonic oscillator.

Working then with $a-b$ we define the \emph{overshoot} $\Omega$ of
an oscillation as
\begin{equation}
  \Omega = \frac{(a-b)_{\textnormal{peak}} - (a-b)_{\textnormal{static}}}{(a-b)_{\textnormal{static}}} \textnormal{.}
\end{equation}
This is motivated again by analogy with the damped mass-spring system, where the overshoot of
the response to a step forcing has a one-to-one correspondence with the damping
ratio \citep[p. 172]{ogata2009}. This measure of the damping is more accurate
here than the standard method of fitting an exponential, since the observed
oscillations have few discernible peaks. Using the overshoot is
less sensitive to uncertainties and can be used even when just one or two peaks
are discernible. Using the formula given by \cite{ogata2009} we compute the damping
ratio $\lambda$ as
\begin{equation}
\lambda = \sqrt{\frac{\ln(\Omega)^2}{\pi^2 + \ln(\Omega)^2}} \textnormal{.}
\end{equation}
Note that it is more traditional in the context of drop oscillations to work
with the damping coefficient $b = \lambda\,\omega_0$, where $\omega_0$ is the natural
frequency, as is done by \citet[p. 474] {lamb1945} and by others. This is a bit
curious, since their results predict a damping ratio which is directly proportional to the Ohnesorge
number, with a proportionality constant depending on the number of the
oscillation mode in question; taking \eg the result by Lamb for the fundamental
oscillation mode of a free droplet we obtain $\lambda = 10/\sqrt{3} \, \times
Oh$. For the damping coefficient $b$ the relationship with $Oh$ also includes $\omega_0$.

Even though it is known that the oscillations of a viscous drop immersed in
a viscous fluid cannot be described by a simple harmonic oscillator
\citep{prosperetti1980}, we posit here that $\lambda$ still gives a good measure of how damped
the drop oscillations are. Note that it follows from the definition that
$\lambda < 1$
corresponds to underdamped oscillations, and that lower values indicate less
damped oscillations. In the Supporting Information we plot the oscillations for
two cases with the same value of $\zeta$ (and thus the same final deformation), but different values of the damping
ratio, together with the step responses of harmonic oscillators with the same
damping ratios. This plot indicates that $\lambda$ is a useful measure of the
damping. 

\begin{figure}
  \begin{center}
    \includegraphics[width=0.8\linewidth]{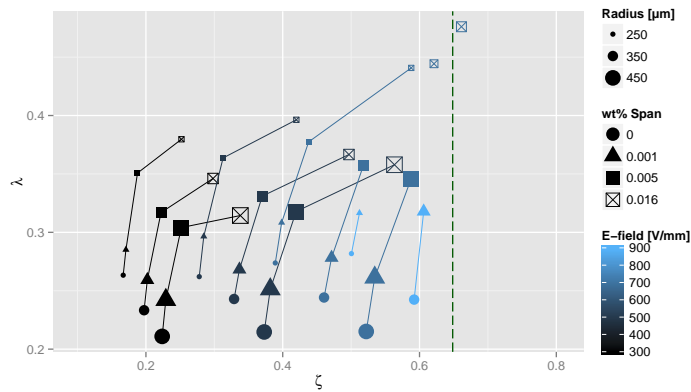}
  \end{center}
  \caption{The damping ratio of oscillations, $\lambda$, versus $\zeta$. Dashed
  line: static deformation limit.}
  \label{fig:sim-taylor-damping}
\end{figure}

Having defined this damping ratio $\lambda$, we show in
\cref{fig:sim-taylor-damping} a plot of $\lambda$ versus $\zeta$ where we connect
points with identical drop size and electric field strength. From this plot we
may surmise that adding small amounts of surfactant increases the damping
significantly, but has a negligible effect on the static deformation
(represented here by $\zeta$), so the lines connecting the 0 wt\% and the
0.001 wt\% results have steep slopes. On the other hand, adding larger amounts
of surfactant significantly increases the $\zeta$ by reducing $\gamma$, so the slopes are flatter.
This can be understood when considering that the surfactants play a dual role in
the system: adding Marangoni stresses and reducing interfacial tension. From
these results we see that when small amounts of surfactant are added, the
increase in damping from Marangoni forces is much more significant than the
reduction in interfacial tension. When more surfactant is added, the effect of
reduced interfacial tension becomes pronounced. We discuss this in more
detail in \cref{sec:discussion}, and illustrate the point with detailed plots 
from the numerical simulations.

Finally, we report the results from a simulation where the drop was falling at
terminal velocity and subjected to a rising voltage pulse (\sparkrise) matching
that used in the experiments. A moving grid procedure was used to keep the falling
drop in the centre of the computational domain. The base value of the applied
field was 500 kV/m, and the fractions 2/7 to 7/7 of this base value were used.

The surfactant concentration was 0.016 wt\% and the drop diameter was 900
$\mu$m.  Based on the terminal velocity $u_T$, the capillary number is $Ca_T =
\eta_2 u_T / \gamma = 0.004$ and the surface P\'eclet number is $Pe_T = D u_T/D_\xi = 8.5$. (Recall that $D$ is
the drop diameter while $D_\xi$ is the surface diffusion coefficient.) For comparison, the electric
capillary number is $Ca_E \in [0.03,0.3]$ for the increasing field strengths. 
We may define an electric surface P\'eclet
number, by analogy with the electric capillary number, as $Pe_E
= D^2\bar{E}^2\epsilon\epsilon_0/(\eta_2 D_\xi)$, which gives in this case 
$Pe_E \in [24,290]$. All in all, these numbers suggest that the external flow is
unimportant, which is also what the simulation results indicate.

\begin{figure}
  \begin{center}
    \includegraphics[width=0.8\linewidth]{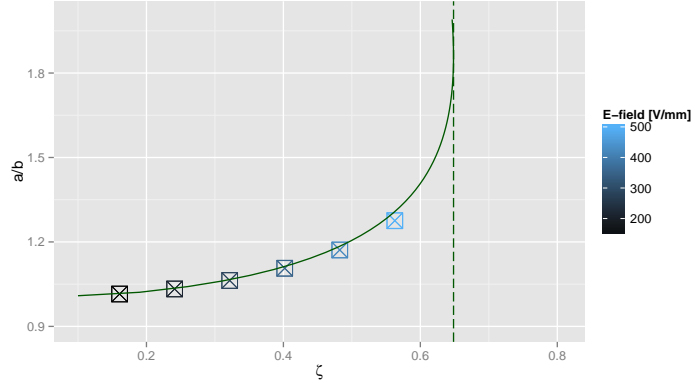}
  \end{center}
  \caption{The static deformation $a/b$ versus $\zeta$ for a 900 $\mu$m diameter
  drop at 0.016 wt\% surfactant concentration. This simulation is with a drop falling under gravity and
subjected to repeated deformations by a rising voltage pulse.}
  \label{fig:sim-taylor-pulsed}
\end{figure}

The simulation results are shown in \cref{fig:sim-taylor-pulsed}. They are very
similar to those seen in \cref{fig:sim-taylor}, which confirms that neither the
external flow nor the previous deformations have a significant influence on the
static deformation in the simulations. If the hysteresis effect observed in the
experiments described in the next section were caused by hydrodynamic effects or
by the surfactant transport, \eg if the flow field was still influenced by the
previous deformation at the start of the next deformation, the hysteresis would
also be observed in these simulations. Since it is not, these can be ruled out
as likely explanations of the hysteresis.

\subsection{Experimental parameter study}
\label{sec:parexp}
Several experiments were performed with different drop diameters, field 
strengths and surfactant concentrations, as described in
\cref{sec:pardesign}. In total, 295 drop deformations were observed, with 8 to
12 observations of each drop and 4 to 6 different voltages applied with both
polarities.

\begin{figure}
  \begin{center}
    \includegraphics[width=0.9\linewidth]{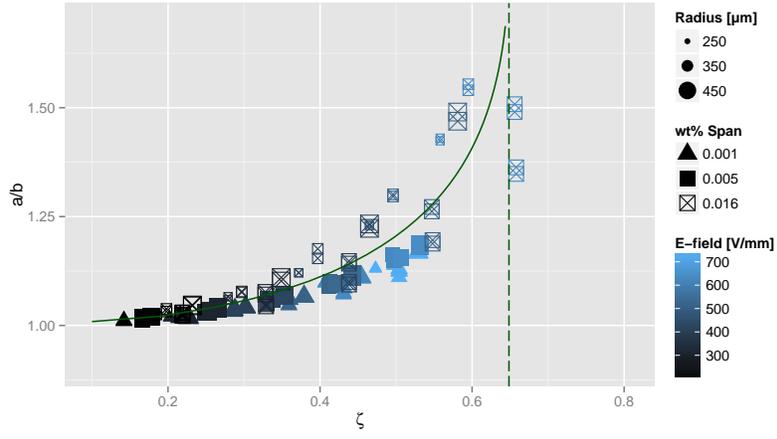}
  \end{center}
  \caption{The deformation $a/b$ found in experiments as a function of $\zeta$
    for \textbf{rising} voltage pulse
  trains. Shaded region: optical measurement
  uncertainty.}
  \label{fig:exp-taylor-rising}
\end{figure}

\begin{figure}
  \begin{center}
    \includegraphics[width=0.9\linewidth]{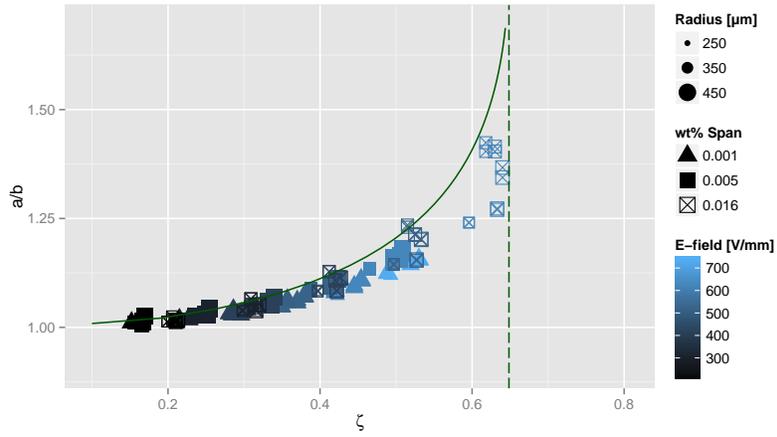}
  \end{center}
  \caption{The deformation $a/b$ found in experiments as a function of $\zeta$
    for \textbf{falling} voltage pulse
  trains. Shaded region: optical measurement
  uncertainty.}
  \label{fig:exp-taylor-falling}
\end{figure}

The results for the static deformation obtained in the experiments are shown in
\cref{fig:exp-taylor-rising} for rising voltage pulse trains (\sparkrise) and
\cref{fig:exp-taylor-falling} for falling voltage  pulse trains (\sparkfall). 
The use of point shapes and sizes match those
used in \cref{fig:sim-taylor} for the simulations.  In these two plots, the shaded
region around the Taylor result indicates the magnitude of the uncertainty
in the optical observations, as described in \cref{sec:expmeth}.

It is seen in these two figures that the experiments are also well-described by
the parameter $\zeta$, and that the results lie fairly close to the Taylor
theory, especially for low deformations. Below $\zeta \sim 0.4$, the deviations
are of the same magnitude as the uncertainty in the optical measurements, while
above this, they are significant. In the Supporting Information we plot also 
the relative deviation $\Delta$ from the Taylor theory is plotted.

There is a profound difference seen between rising and falling pulse trains in
these plots, in that the deviation is positive for the latter, but both
positive and negative for the former.
Furthermore it is seen that this difference occurs only for the highest
concentration of surfactants. 

\citet{peltonen2000} have
reported similar hysteresis effects when repeatedly stretching and
compressing an interface
between water and hexane with added Span 80, using
a Langmuir-Blodgett apparatus. 
When comparing \cref{fig:exp-taylor-rising}
and \cref{fig:exp-taylor-falling} given here, it is noted that for rising
voltage pulses, the previous stretchings at small and intermediate field
strengths significantly affect the subsequent stretchings, giving a deviation
from the Taylor theory that has the opposite sign of that seen in all other
cases. Note also that the simulation results (\cref{fig:sim-taylor}) show
positive deviations for all but one point.

The hypothesis by \cite{peltonen2000} is that earlier compressions
disperse surfactants into the water phase, but it is not readily apparent that
this is the case here; if this were so, the large expansions and compressions
of the interface
that occur at the beginning of a falling voltage pulse train should
significantly affect the subsequent medium and small expansions, but this is
not observed.

For the sake of clarity, we remark that the hysteresis seen here is an entirely different
phenomenon from the hysteresis studied \eg by \citet{sherwood1988}. Sherwood
considers the hysteresis in the deformation $a/b$ which arises from a finite
permittivity ratio, but as is apparent from his figure 2, the phenomenon he
discusses requires deformations $a/b \gg 10$, while our deformations are all $a/b
< 2$. Thus the hysteresis phenomenon observed herein cannot be attributed to
a finite permittivity ratio.

\subsection{Comparison of simulations and experiments}
\label{sec:discussion}
Several features of the results of these studies warrant further comment. First
of all, we observe stable solutions slightly beyond the limit predicted by
Taylor. This is attributable to the interfacial elasticity. 
At the highest surfactant concentration and the largest
drop deformations this results in a decrease of about 2\% in $\zeta$, which is 
sufficient to explain the deviation. However, for the lower surfactant
concentrations the effect is too small to account for the observed deviation.
It is likely that this discrepancy is caused by some of the approximations
used by Taylor. Other authors have also pointed out minor disagreements between
their results and the Taylor theory. \citet{brazier1971}, using numerical
iterations to obtain pressure balance along the entire interface, found a difference
which is very similar to that found here.

As for the general agreement with the Taylor theory, it is apparent from the
numerical results that the main effect of surfactants on the static
deformation is through the reduction in equilibrium surface tension. When this
is taken into account in the calculation of $\zeta$, the results agree nicely
with Taylor's prediction, as shown in \cref{fig:sim-taylor}.

For the experimental results, however, there is a clear tendency for smaller
deformations than those predicted by the Taylor theory. One possibility is that
this discrepancy is caused by a form of interfacial elasticity which is not
accounted for here, as discussed earlier.

\begin{figure}
  \begin{center}
   \begin{subfigure}[t]{0.45\linewidth}
    \includegraphics[width=\linewidth]{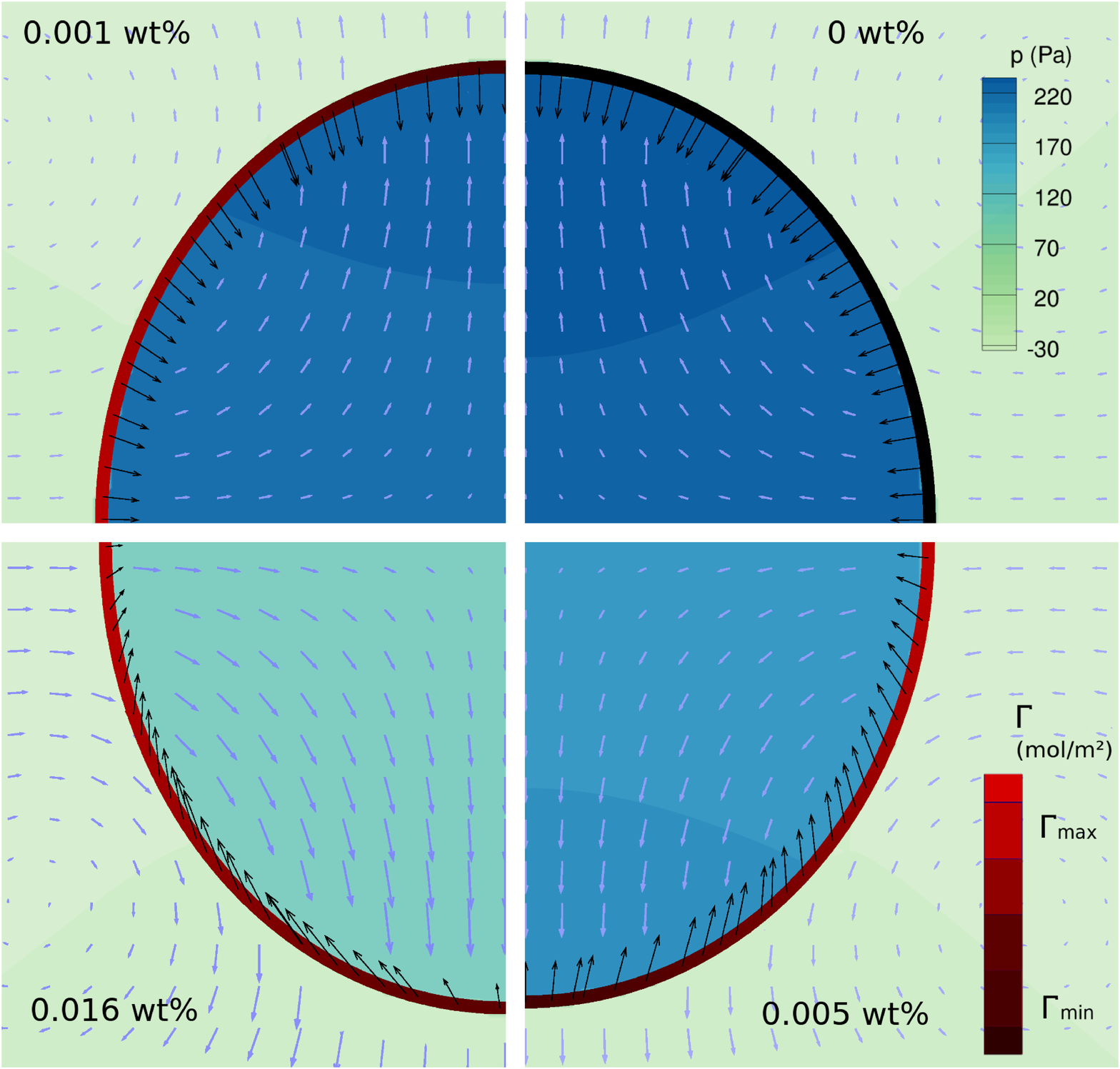}
    \caption{Comparison of the pressure field (blue/green), the vector
      $\gamma\kappa\v{n} - 100 \grad_{\iota}\gamma$ (black vectors), the
      flow field (sky-blue vectors) and the surfactant concentration at the
      interface (red to black corresponding to the variation shown in
      \cref{fig:time-evolution-surfactant}, black for 0 wt\%
      corresponding to $\Gamma = 0$). The quadrants show the four
      different bulk concentrations considered in this paper.}
   \label{fig:sim-surften-marangoni-a}
   \end{subfigure}\hfill
   \begin{subfigure}[t]{0.45\linewidth}
    \includegraphics[width=\linewidth]{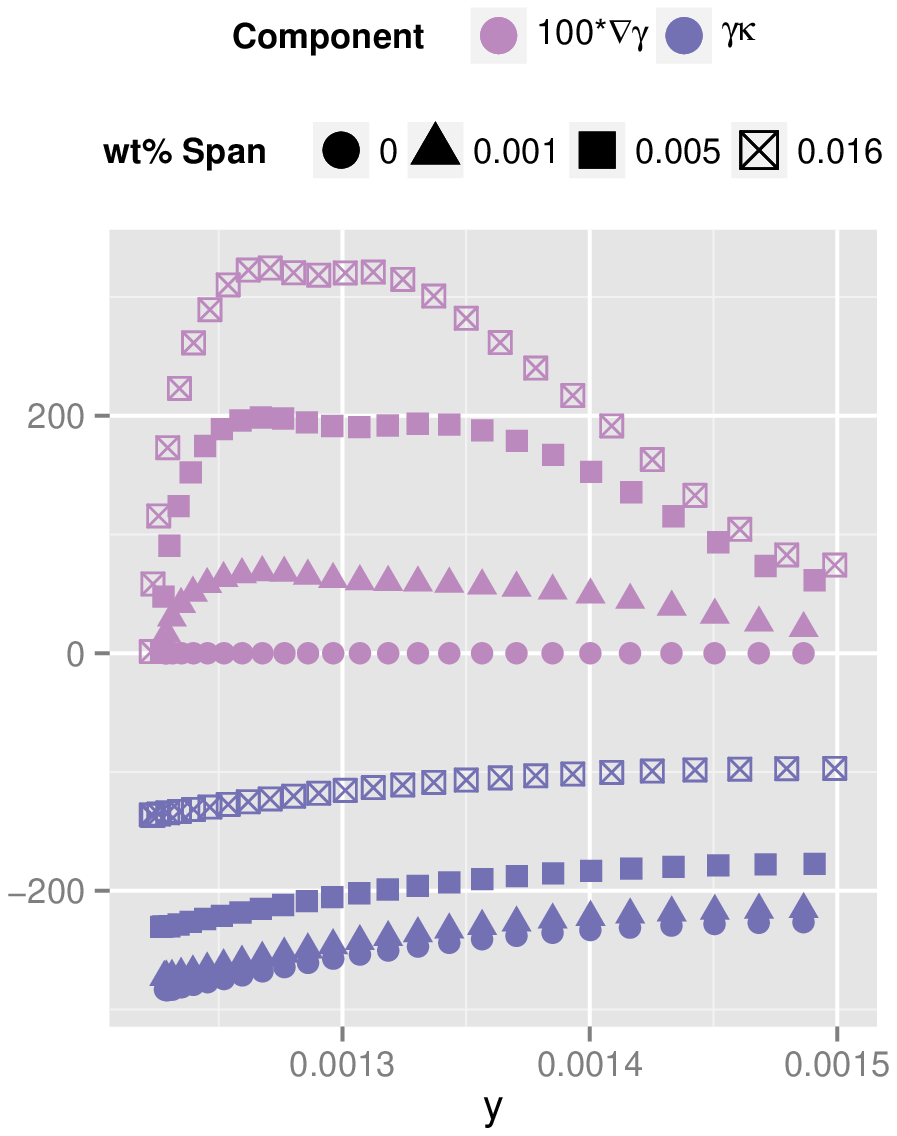}
    \caption{The quantities $\gamma \kappa$ and
      $100|\grad_{\iota}\gamma|$ along the interface. The abscissa here
    is the vertical coordinate in (a), such that the left end of this plot corresponds to the
  pole and the right end to the equator of the drop.}
    \label{fig:sim-surften-marangoni-line}
   \end{subfigure}
  \end{center}
  \caption{The effect of surfactant concentration on normal and tangential
    interfacial stress. The plots are for the
  0.5 mm diameter drop subjected to a 700 V/mm electric field, at $t=1$ ms
  corresponding to the blue vertical line in
  \cref{fig:time-evolution-surfactant}. The values of
  $\zeta$ are 0.39, 0.40, 0.44 and 0.59 in order of increasing surfactant
concentration.}
  \label{fig:sim-surften-marangoni}
\end{figure}

As illustrated in \cref{fig:exp-taylor-damping-comparison}, the effect
of adding surfactant is two-fold. Small additions increase the damping
significantly, while the reduction in interfacial tension is small, so the
static deformation is not much affected. Conversely, adding larger amounts
of surfactants give a significant decrease in interfacial tension causing
a larger deformation, while the increase in damping is less significant.
In order to explain this effect, we show in \cref{fig:sim-surften-marangoni-a}
plots of the pressure field as well as the vector quantity $\gamma \kappa \v{n}
- 100 \times \grad_{\iota} \gamma$
at the interface for the four different surfactant concentrations
considered here. This is for the 0.5 mm diameter drop subjected to a 700
V/mm electric field. We scale the Marangoni force by 100 to accommodate the
visualisation, since the
curvature $\kappa$ is very large for these small drops. The important thing
here is not the absolute value of this vector, but rather the comparison
between the four cases in the tangential and normal components. All plots are
shown at the same time, $t=1$ms,
which is a little less than halfway to the peak deformation. Around this time,
the Marangoni forces are at their largest, since these forces counteract
the surfactant maldistribution driving the concentration profile to be uniform
at equilibrium. This time is also convenient since the deformation of the drops
is very similar at this point in time, while they differ more at later times. 
Also shown in this plot is the flow field, and the surfactant distribution
along the interface in red colour going from the lowest (darkest) to the
highest (brightest) concentration along the interface in each case. The interfacial
positions are very similar at this early time, which in turn means
that both the electric
fields and the curvature profiles are also very similar. However, note that the
flow is stronger for the drop with the highest surfactant concentration, consistent with the
fact that this will be more deformed than the other drops.

\begin{figure}
  \begin{center}
    \includegraphics[width=0.9\linewidth]{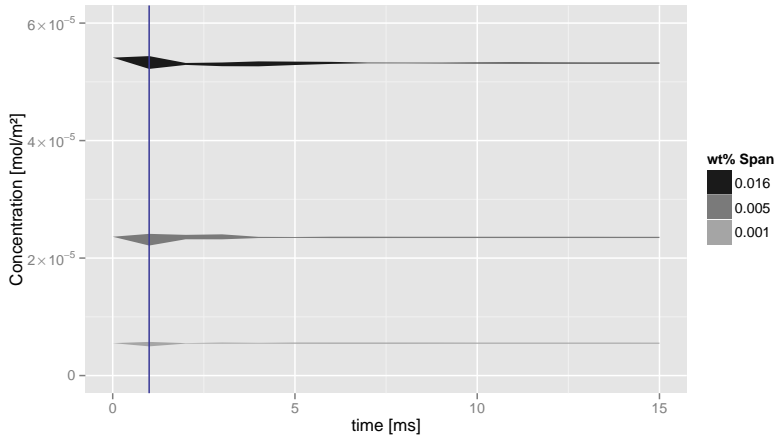}
  \end{center}
  \caption{The time evolution of the minimum and maximum interfacial
  surfactant concentration, shown as the lower and upper edges of shaded bands,
  for the three cases with surfactants shown in
  \cref{fig:sim-surften-marangoni}. The vertical blue line shows the
  instant at which \cref{fig:sim-surften-marangoni} is plotted. Note that
all values are well below $\Gamma_\infty = 1.31 \times 10^{-4}$ mol/m$^2$, \ie
within the range of validity of the surfactant EoS.}
  \label{fig:time-evolution-surfactant}
\end{figure}

\begin{figure}
  \begin{center}
    \includegraphics[width=0.8\linewidth]{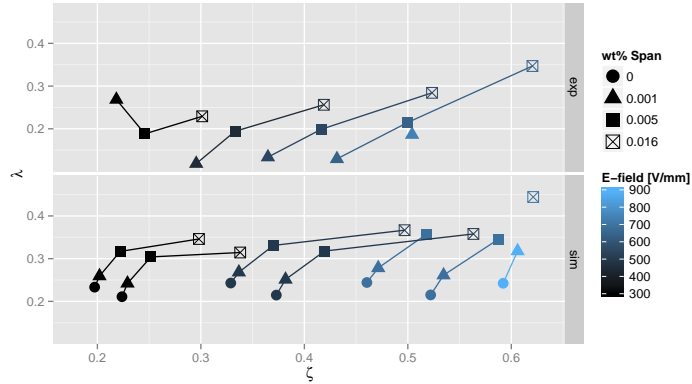}
  \end{center}
  \caption{The damping ratio $\lambda$ versus $\zeta$, for the experimental
values (top) and the simulation results shown in
\cref{fig:sim-taylor-damping} (bottom). The simulation results are only shown for drops
with $D\ge 600 \mu$m to avoid clutter.}
  \label{fig:exp-taylor-damping-comparison}
\end{figure}

In all the three cases where surfactants are present, it is
seen that the initial deformation gives an increased surfactant concentration near the
equator and a reduced concentration near the poles. The lowest
and highest concentrations $\Gamma_{\textnormal{min}}$ and $\Gamma_{\textnormal{max}}$ for each
case with surfactant present are plotted in \cref{fig:time-evolution-surfactant}
as functions of time. 
For comparison, the maximum
possible interfacial concentration given by fitting the Langmuir EoS
\cref{eq:langmuir} to the experimental data is 1.31 $\times 10^{-4}$ mol/m$^2$,
and all values are well below this.

In \cref{fig:sim-surften-marangoni-line} we show the two components of the
vector $\gamma \kappa \v{n} - 100 \cdot \grad_{\iota} \gamma$ plotted as a
function of the vertical coordinate $y$, covering here one quadrant of the drop. Inspecting the plot, it is seen that the Marangoni
forces increase significantly at the lower surfactant concentrations, while the
decrease in interfacial tension is significant mainly for the highest surfactant
concentration. These plots
confirm the hypothesis put forward to explain the influence of surfactants on
the damping which is seen in \cref{fig:sim-taylor-damping}.

Another interesting phenomenon seen in the numerical results is the effect of
surfactant concentration on the damping of oscillations, cf.\
\cref{fig:sim-taylor-damping}. Naturally,
it is interesting to see if the experiments show a similar trend. Since the experimental
results do not admit the same direct comparison by holding two parameters
identical while varying a third, the electric field was binned into 5 intervals with
limits at $(200,400,500,600,700,800)$ V/mm. No experiments were done
with electric fields below 200 or above 800 V/mm. We then considered all results
within such a bin and with a given surfactant concentration and joined these
into groups, amounting to averaging over the radius. This gave $5\times3=15$ groups. 
We omit deformations where the difference between the maximum and the
static deformation was smaller than the optical resolution, \ie we only consider
observably underdamped oscillations.

To create a plot like \cref{fig:sim-taylor-damping}, the centre-of-mass of each
group was computed, connecting centre-of-mass points that represent groups with
the same range of electric fields.  This corresponds to averaging over the drop
radius. Plotting this together with
the simulation results in \cref{fig:exp-taylor-damping-comparison}, it is seen
that a similar trend is found, in particular for the slopes between the 0.001
and 0.005 wt\% results. However, the absolute value of the damping is lower 

In this comparison, note that the
circular points corresponding to zero surfactant can only be shown for the
simulation results, since the system is known to be contaminated even when no Span 80 is added. 
Note also that in the plot of the simulation results, only
those drops with $D\ge 600\mu$m are shown in order to reduce clutter.

When it comes to the hysteresis, we have no clear explanation of the
observed phenomena. It is evident from the results shown in
\cref{fig:sim-taylor-pulsed} that the simulations cannot explain the
hysteresis, even when taking into account the external
flow, the history of previous deformations and the Marangoni effect. There
is a possibility that incorporating the elastic effects caused by the
surfactant could help explain the hysteresis, but this does not seem to be likely
\emph{a priori}, since the forces from the elasticity are
expected to decrease the deformation, not increase it. Our results do not appear to support the
hypothesis by \cite{peltonen2000}, where it is proposed that the hysteresis can be
explained by the previous
deformations causing surfactants to detach into the water phase. If this
were the case, we should observe the hysteresis also for falling voltage
pulses, and we do not. If one may speculate, it could be that the small deformations caused
by the initial part of a rising pulse train can cause a phase transition in the
monolayer of surfactants at the interface. Such phenomena have been
reported in the literature \citep{ravera2005}.

\section{Concluding remarks}
We have performed detailed studies of the effect
of surfactants on the electrohydrodynamic stretching of water drops in oil at
various drop sizes and electric field strengths, covering the full range of dimensionless
electric field $\zeta$ from zero to drop breakup. We
have compared our results to the classic result by Taylor, which assumes no
surfactants present at the drop interface, and predicts the deformation as
a function of $\zeta$.

We find that when the equilibrium interfacial tension caused by the surfactant
is used in the expression for $\zeta$, the system remains well-described by this
dimensionless quantity, as expected from \cref{fig:time-evolution-surfactant}
which shows that the surfactant maldistribution quickly becomes small. For field
strengths below $\zeta \sim 0.4$, \ie deformations below $a/b \sim 1.12$, we
have found only negligible deviations from the Taylor theory when surfactants
are added. For field strengths above this we have reported significant
deviations of the observed drop deformations, as well as an ability to go
slightly beyond the critical stability limit, $\zeta \sim 0.65$, predicted by
the Taylor theory without drops breaking up. The deviations from the Taylor
theory are larger for experimental than for the simulation results, which could
be explained by the interfacial elasticity caused by the surfactant, an effect
which is not taken into account in the simulations.

We have shown both by simulations
and experiments that the addition of surfactants damps the oscillations
induced by a suddenly applied electric field, an
effect which may be attributed to the Marangoni effect that arises in the
presence of interfacial-tension gradients. 
We have studied the effects of the
surfactant concentration on the damping of oscillations, and found that low
concentrations increase the damping significantly while having little effect on
the static deformation. On the other hand, the difference between low and high
surfactant concentrations lies mainly in the change of equilibrium surface
tension, which affects the static deformation.

Finally we have observed in our experiments
a significant hysteresis effect when repeatedly stretching drops at high
surfactant concentrations. But these effects are only seen for
the case where the applied deformations are first small and then increased, not
in the opposite case when the deformations are first large and then decreased.
We have investigated whether this effect can be explained just by the
hydrodynamics and surfactant transport, and have found this not to be the case.
One may speculate that the small initial deformations give the surfactant heads
enough room to reorient into an energetically more favorable state, thus making
the interface more pliable.


The data produced by the parameter studies in this paper are permanently stored
at Figshare, \url{http://dx.doi.org/10.6084/m9.figshare.1254343}.

\section*{Acknowledgments}
We would like to thank Dr.\ Martin Fossen (SINTEF Petroleum Research) for the
measurements of interfacial tension as a function of surfactant concentration,
Dr.\ Velaug Myrseth Oltedal (SINTEF Petroleum Research) for the measurements of
bulk viscosity, and Dr.\ Cédric Lesaint (SINTEF Energy Research) for the
measurements of density and for enlightening discussions of the interfacial tension
measurements. We would also like to thank Dr. Gunnar Berg (SINTEF
Energy Research) as well as Professor Jean-Luc Reboud and Dr. Pierre Atten (G2Elab)
and Dr. Erik Bjørklund (Sulzer Chemtech)
for fruitful discussions on the work presented here.

This work was funded by the project \emph{Fundamental understanding of
electrocoalescence in heavy crude oils} coordinated by SINTEF Energy Research.
The authors acknowledge the financial support from the Petromaks programme of the
Research Council of Norway (206976), Petrobras, Statoil and Sulzer Chemtech.

\bibliographystyle{abbrvnat}
\setcitestyle{authoryear,open={(},close={)}}
\bibliography{references}

\end{document}